# One-step fabrication of flexible, cost/time effective and high energy storage graphene based supercapacitor

Gholami laelabadi, Katayoon[1,2]; Moradian, Rostam[1,2]*; Manouchehri, Iraj[1,2]

[1]Physics Department, Faculty of Science Razi University, Kermanshah, Iran

[2]Nano science and nano technology research center, Razi University, Kermanshah, Iran

The advances in micro-size and in-plane supercapacitors lead to produce the miniaturizing energy storage devices in portable and bendable electronics. Micro-supercapacitors have the unique electrochemical performance, such as high power density, fast charging, long cycle life and high safety. The reduction time and cost in fabrication processes of micro supercapacitors are important factors in micro-fabrication technology. In this work, a simple, scalable and cost-effective fabrication of interdigitated reduced graphene oxide@polyaniline flexible micro supercapacitors is presented. We found that in fabricating the interdigitated micro electrode patterns on PET substrate; the reduction of graphene oxide and growth of conducting polymer are rapidly performed simultaneously in one step by laser irradiation. The capacitance was 72 mF/cm$^2$ at 35μA/cm$^2$ current density. These highly capacitance micro supercapacitors demonstrate good stability and more than 93.5% of the capacitance retain after 1000 cycles at 0.7 mA/cm$^2$ current density.

Nowadays, design and fabrication of energy storage systems is essential for saving clean and high efficiency energy sources such as solar energy. Development of portable and flexible electronic devices including flexible displays[1], curved smart mobile phones[2], robotic skins[3] and implantable medical devices[4], have attracted numerous studies to investigate light, thin, eco-friendly and flexible supercapacitors. Flexible micro-supercapacitors (MSCs) with ultra-high power density, large storage charge and superior cycling lifetime are offered as a stage of next production on-chip energy storage devices. One of the challenges to improve energy storage properties of supercapacitors is the development of suitable electrode materials with good mechanical flexibility, high energy density and excellent electrochemical stability. Recently, graphene has attracted intensive attentions for manufacturing planar MSCs[5,6]. Formation of electrochemical double layer capacitors on the surface of graphene sheets; lead to large specific surface area, high electrical conductivity and excellent chemical and electrochemical stability for electrode materials of MSCs[5,6]. Based on the mechanism of energy storage, supercapacitors can be divided into two main categories: electric double-layer capacitors and pseudocapacitors. The former usually involves carbon-based materials which charge storage process performs at the electrode/electrolyte interface while in the latter, energy stores through faradic reactions in conductive polymers or metal oxides electrode materials. The faradic reactions take place during the charge/discharge processes, hence pseudocapacitors usually improve electrochemical performance. The supercapacitors based on composite materials of graphene and numerous conductive polymers have been reported to achieve high capacitance[7]. The conducting polymers including polyaniline (PANI), polypyrrole (PPy) and poly (3,4-ethylenedioxythiophene) (PEDOT) have facile synthesis, low cost, high flexibility and high pseudocapacitance[8]. Polyaniline synthesis does not require any special equipment or precautions and its electrochemical behavior shows multi-redox state so reduced graphene oxide/polyaniline (RGO@PANI) composites have attracted significant attentions. In addition, PANI is the only conducting polymer whose electronic structure and electrical properties can reversibly be controlled by both oxidation and protonation[9] and the high conductivity and synergistic π-π effect from interaction between the aromatic rings of PANI and graphene makes it useful as supercapacitor electrodes. Generally, PANI and grapheme construct a composite by chemical or electrochemical methods via two steps: 1-polymerization of aniline and 2-dispersion of PANI homogeneously onto the surface of graphene sheets which are prepared by reduction of graphene oxide. For example, Jianglin Ye *et al.* obtained monolayer graphene films on Cu substrate by CVD method then deposited aniline through electropolymerization on the graphene films[10]. Finally, they fabricated interdigital micro electrode by direct laser writing. In their method since they have just one layer of graphene coated surface for depositing aniline therefore, capacitance/area of their supercapacitor device is low. Most of works are based on graphene oxide (GO) as precursors; Zhang *et al.* mixed aniline monomer with graphene oxide sheets in aqueous solution, after polymerization and formation of RGO@PANI composite, it reduced by hydrazine at 98 °C for one day[11]. They studied electrochemical behaviors of graphene/polyaniline electrodes in a three-electrode configuration cell. However, fabricating uniform layers of RGO@PANI composite with suitable porous structures as advanced electrochemical electrode materials remains a great challenge. In this work, we demonstrate a novel, cost/time effective, scalable, facile and efficient approach to fabricate flexible planar micro supercapasitor based on graphene/polyaniline composite films on polyethylene terephthalate (PET) substrates. The composite material electrodes prepared in one step and fabrication of interdigital pattern and reduction of graphene oxide/polyaniline (GO@PANI) composite performed simultaneously by laser beam irradiation. To investigate advantages of conducting polymer on RGO sheets, we fabricated an interdigitated micro-supercapacitor by laser's photothermic reducing effect on graphene oxide. The presence of polyaniline on RGO sheets leads to increase stack areal capacitance about 13 times higher than that of RGO.

**Results**

**Fabrication of laser reduced graphene oxide@polyaniline micro supercapacitors.** Fig. 1 schematically shows a simple process to obtain interdigitated laser reduced graphene oxide@polyaniline (LRGO@PANI) micro electrodes. The pattern of Au current collectors are deposited by shadow masking thermal evaporation method on flexible PET substrates. Then, GO@PANI solution is drop casted on it. The laser beam irradiation forms LRGO@PANI micro patterns. The photograph of the as-prepared interdigitated micro electrodes are shown in Fig. 1 for each step. For comparison, the interdigitated laser reduced graphene oxide (LRGO) micro electrodes are prepared using the same method and graphene oxide (GO) solution. This form of planar micro supercapacitors based on LRGO@PANI and LRGO micro electrodes are called LRGO@PANI-MSC and LRGO-MSC, respectively.

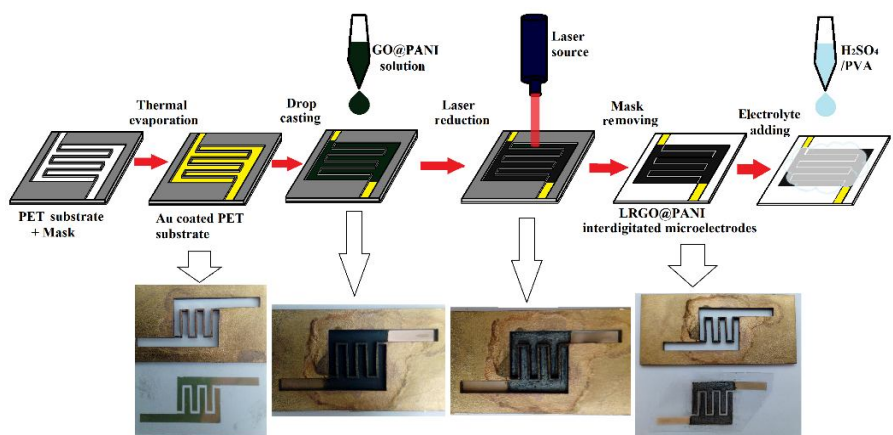

**Fig. 1: Schematic and digital photograph of the fabrication steps for LRGO@PANI-MSC.**

**Morphological and structural study.** Fig. 2 shows FE-SEM of LRGO and LRGO@PANI morphologies. Fig. 2(a1-a5) and Fig. 2(b1-b5) show the FE-SEM images of GO and GO@PANI nanocomposite reduced by laser at different scales, respectively. It is observed, that similar to the previous reports[12,13], the ablated porous structure with continuous interconnected framework are formed in all samples. This is because laser beam provides localized high temperature that causes chemical bond cleavage and sublimation of atoms to recombine into gaseous products, which lead to hydrodynamic drainage and porous structures[12]. Presence of polyanilline in RGO sheets, forms densely packed nanoparticle clusters in the form of coral-like structures (comparing images in Fig. 2(a1), (b1). It is observed that LRGO@PANI nanocomposite is more porous and disordered than the LRGO. Porosity of supercapacitor electrode is an important parameter to facilitate ions diffusion in the charge/discharge process which leads to better contact with the electrolyte and providing numerous channels for ion transportation and diffusion. Therefore, presence of polyaniline in the structure of LRGO@PANI, not only improve conductivity and flexibility of the provided electrode, but also increases its cycling stability via preventing the volume change and particle aggregation during the charge/discharge process[14].

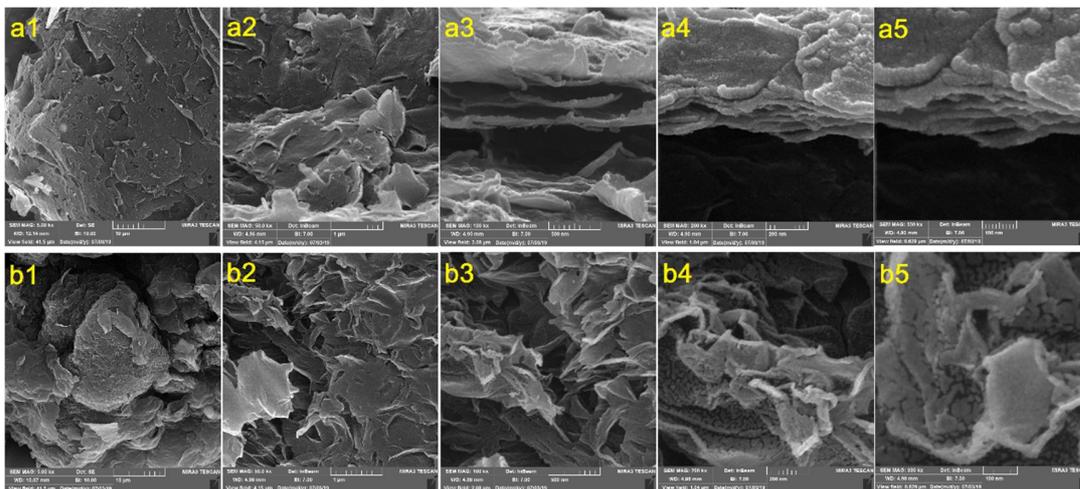

**Fig. 2: FE-SEM images of LRGO and LRGO@PANI electrodes.** FE-SEM images at different scales (100nm- 1μm) show direct reduction and expansion of (**a1-a5**) GO and (**b1-b5**) GO@PANI by laser beam.

The growth of multilayers of RGO and RGO@PANI nanocopmosite sheets from Fig. 2(a5), (b5) indicate an interconnected porous structure with nanometer-scale layers size. It is found that average thickness of the multilayers is about 20 nm for LRGO@PANI sample which is quite small compared to LRGO sample (see Supplementary Fig. S1). Small nanosheets provide higher electroactive regions and shorter diffusion paths, which help to improve the electrochemical applications of LRGO@PANI nanocomposite electrodes. The PANI chains are not observable in LRGO@PANI because the RGO nanosheets are distinctly wrapped the PANI nanostratures due to different possible interactions such as electrostatic interactions, π-π stacking and hydrogen bonding[15].

The typical FE-SEM images of finger electrodes and the photographic images of the fabricated LRGO@PANI micro-electrodes, which is prepared by mask shadowing laser beam, is demonstrated in Fig. 3a-c. The top view FE-SEM images of finger electrodes in Fig. 3(a1), (b1) show that the obtained large-area, millimeter-long and well-aligned LRGO and LRGO@PANI interdigital fingers on PET substrate, respectively. The inset images of the finger electrode are illustrated in the Fig 3a, b. The width of the finger electrodes and the interspaces are about ~ 2 and 1 mm, respectively. Although a few isolated active materials in small pieces is observable in the interspace but the interspace lines serve as good separators between the positive and negative interdigitated electrodes. There are protuberances in finger electrodes in Fig. 3a, b, which come from the expansion of the film and the aggregation of reduced graphene oxide flakes when treated with the laser, thus enabling full access to the electrode surface that is essential for charging the electrodes. To measure finger electrodes' thickness, the edge of one of the finger electrode is characterized by the sectional FE-SEM image. The cross-sections images of the micro-electrode of LRGO (Fig. 3(a2), inset) and LRGO@PANI (Fig.

3(b2), inset) reveal thickness of ~70 μm. Furthermore, it is observed that adhesion of LRGO@PANI electrode's active material to the substrate is much better than that of LRGO electrodes. Fig. 3(c1) shows the photographic image of mask and interdigitated electrodes of LRGO@PANI and flexibility of this micro electrodes, Fig. 3(c2). There is no short circuit between two electrodes in the pattern of the interdigitated electrodes.

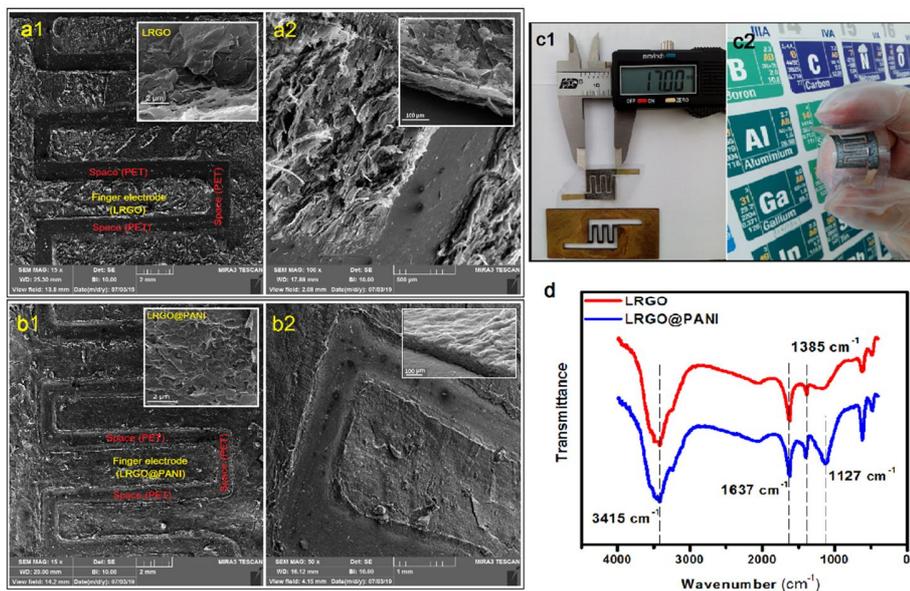

**Fig. 3: Characterization of microelectrodes.** The top view FE-SEM images of (**a1, 2**) LRGO and (**b1, 2**) LRGO@PANI finger electrodes show large-area and good reduction process. (**c**) The photographic images of mask and LRGO@PANI flexible microelectrodes. (**d**) FTIR spectra of LRGO and LRGO@PANI microelectrodes.

Fig. 3d shows Fourier transform infrared (FTIR) spectra for LRGO and LRGO@PANI films. The intense peaks in the FTIR spectrum of the LRGO reveal evidence of stretching vibrations: O–H absorption causes the peak at 3415 cm$^{-1}$ and peak at 1637 cm$^{-1}$ is due to in-plane C=C bonds and the skeletal vibration of the graphene sheets. The weak peak at 1385 cm$^{-1}$ in the FTIR spectra corresponds to vibrations of C–OH bonds of graphene-hydroxyl groups. As reported in the literature, remnants of hydroxyl groups related to water molecules intercalating among LRGO layers, do not affect the hydrophobicity of the LRGO and usually improve its electrical conductivity[16]. FTIR spectra of LRGO@PANI film provided evidence of interaction between graphene sheets and PANI chains in the nanocomposite. The spectrum of LRGO@PANI exhibited almost the same vibrational bonds as LRGO sample, but peak at 1127 cm$^{-1}$ is appeared in nanocomposite electrode. This peak is assigned to the linear PANI backbone and C–N stretching of the quinoid ring which shows covalent interaction between the π bonded structures of the conjugated PANI grafted on RGO sheets. It is described as the electronic-like absorption peak characteristic of conducting PANI and exhibits the extent of electron delocalization[17,18]. Because of the π–π interaction between the graphene sheets and PANI in the composite in LRGO@PANI sample, the intensity and wavenumber of C–OH bonding peak are higher than that of LRGO sample (the wavenumber changes from 1385 to 1402 cm$^{-1}$) which lead to limitation of C-OH vibrations.

Raman spectroscopy is a powerful tool for characterizing graphene flakes. It is commonly used to determine the number of graphene layers, the defect density, edge chirality and strain. The Raman spectra of GO, LRGO and GO@PANI, LRGO@PANI are shown in Fig. 4. For all samples, the two most intensive peaks occur in the ranges 1351-1361 cm$^{-1}$ and 1588-1593 cm$^{-1}$. The Raman band at 1351-1361 cm$^{-1}$ and 1588-1593 cm$^{-1}$ known as the D and G bands, respectively. G band is a doubly-degenerate in-plane sp$^2$ C–C stretching mode that originates from phonons at the Γ point in the center of the first Brillouin zone. This band exists for all sp$^2$ carbon systems due to phonons emitted by excited electrons or the holes which are scattered by incident photons[19]. D band originates from a defect that includes any breaking of the symmetry of graphene lattice and any changing in carbon hybridization such as $sp^3$-defects, grain boundaries or even an edge which can be introduced by interfacing at the borders of crystalline areas[20], for momentum conservation. In this band the electron is elastically scattered by a phonon to the K' point and then elastically back-scattered to the K point by a defect. The ratio of D and G bands peak intensities ($I_D/I_G$) can be used experimentally to determine the extent of defects on GO and RGO films. Graphene oxide is a compound of carbon, oxygen and hydrogen in variable ratios, Fig. 4a, b indicate the strong D band and $I_D/I_G \geq 1$ in GO and GO@PANI samples that confirm its lattice distortions and a large amount of $sp^3$-like defects caused by the oxidation process and the presence of polymer in the structure of GO (in GO@PANI films). While the intensity ratios of $I_D/I_G$ of LRGO and LRGO@PANI samples are less than that of GO and GO@PANI. It is because after reduction process by laser light, the intensity of G band increases which corresponds to the recovery of the hexagonal network of carbon atoms, therefore the number of the graphene-like domains is increasing.

Raman spectrum of GO@PANI and LRGO@PANI composites, Fig. 4b, show that the signals for @PANI cannot be clearly observed in the composite because the PANI chains have been enveloped by the graphene and graphene oxide sheets. This illustrates that PANI in the samples are well wrapped in the layer of graphene (and graphene oxide) sheets, in accordance with FE-SEM results Fig. 2 which is in agreement with the report of Yongsong Luo *et al*[21]. Comparison of GO and GO@PANI Raman spectra is shown in Fig. 4c. The frequencies of D and G peaks in GO@PANI nanocomposite film, blue shifts about ~5 cm$^{-1}$ from GO film. The peak positions are sensitive to external perturbations, in GO@PANI film the growth of polymer chains on graphene oxide sheets changes band structure; hence carrier concentration; the formation of a low defect concentration and charge transfer from defects[22]. The lower thickness of multilayers in GO@PANI with respect to GO film leads to more quantum confinement effect hence a blue shift in Raman spectra.

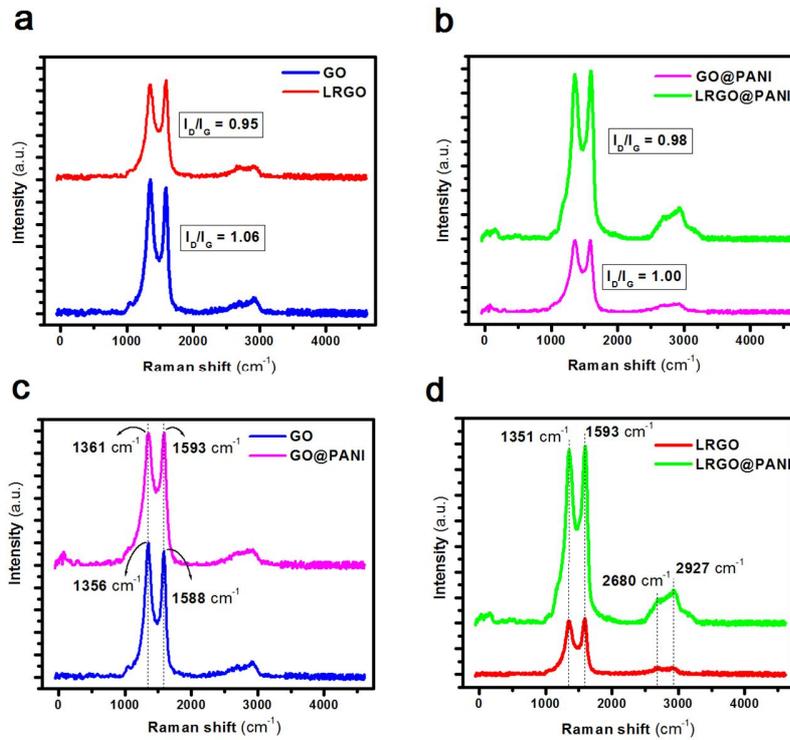

**Fig. 4: Raman spectrum of microelectrodes.** A comparison of Raman spectrum for (**a**) GO and LRGO, (**b**) GO@PANI and LRGO@PANI, (**c**) GO and GO@PANI, (**d**) LRGO and LRGO@PANI microelectrodes.

Fig. 4d indicates Raman spectra of LRGO and LRGO@PANI films. The weak peaks at 2680 cm$^{-1}$ and 2927 cm$^{-1}$ correspond to 2D and D+G band, respectively[19]. 2D band is a second order Raman process originating from the in-plane breathing-like mode of the carbon rings. In double resonance process, an electron–hole pair is created by an incident photon near the K point in the first Brillouin zone. 2D peak shows characteristic of sp$^2$ hybridized carbon‑carbon bonds in graphene, that for single layer it is very sharp however by increasing number of graphene layers, the peak intensity reduces and broadens; 2D and D+G peaks in the as prepared films show presence of multilayer graphene, as seen in FE-SEM images (Fig. 2). In Fig. 4d the peaks positions do not change with the presence of conductive polymer however the intensity increase, which may be due to the change in structure by increasing functional group and density, hence vibrational modes increase. The recovery of the 2D peak is clearly observed in LRGO@PANI film. Since 2D band is more sensitive to the defects of graphene-based materials, the regular sp$^2$ bonding network is dominated in the presence of polyaniline in the structure of the RGO films.

**Electrochemical measurements of MSCs.** To evaluate the supercapacitive performance of the fabricated flexible micro electrodes, cyclic voltammetry (CV), electrochemical impedance spectroscopy (EIS) and galvanostatic charge/discharge (GCD) tests were carried out according to Stoller & Ruoff recommendations[23] and others[12,24] in a two-electrode system (packed micro-supercapacitor). The CV curves of LRGO-MSC and LRGO@PANI-MSC at various scan rates (5, 10 and 20 mV/s) are shown in Fig. 5a, b, respectively. In Fig. 5a, within the employed potential range (from -1.0 to 1.0 V), the curves exhibit nearly rectangular shapes without chemical reaction peaks that could be associated with Faradic effect therefore, suggesting formation of an effective double electric layer capacitance in the LRGO micro electrodes. Here, a hydrogel polymer electrolyte, polyvinyl alcohol (PVA)/$H_2SO_4$, is used to fabricate the all-solid-state MSCs. The CV curves of micro electrodes based on graphene@polyanilline nanocomposites, Fig. 5b, show a quasi-rectangular shape with a pair of redox peaks (~0.3/-0.3 V and ~0.7/-0.7 V), which can be ascribed to reversible charge-discharge behavior and pseudocapacitance characteristic of polyaniline due to a comprehensive effect of the changing in polyaniline structures. The pair of peaks 0.3 & -0.3 V is attributed to the redox transition of PANI between leucoemeraldine and meraldine. Another pair of peaks 0.7/-0.7 was derived from the emeraldine-pernigraniline transformation.

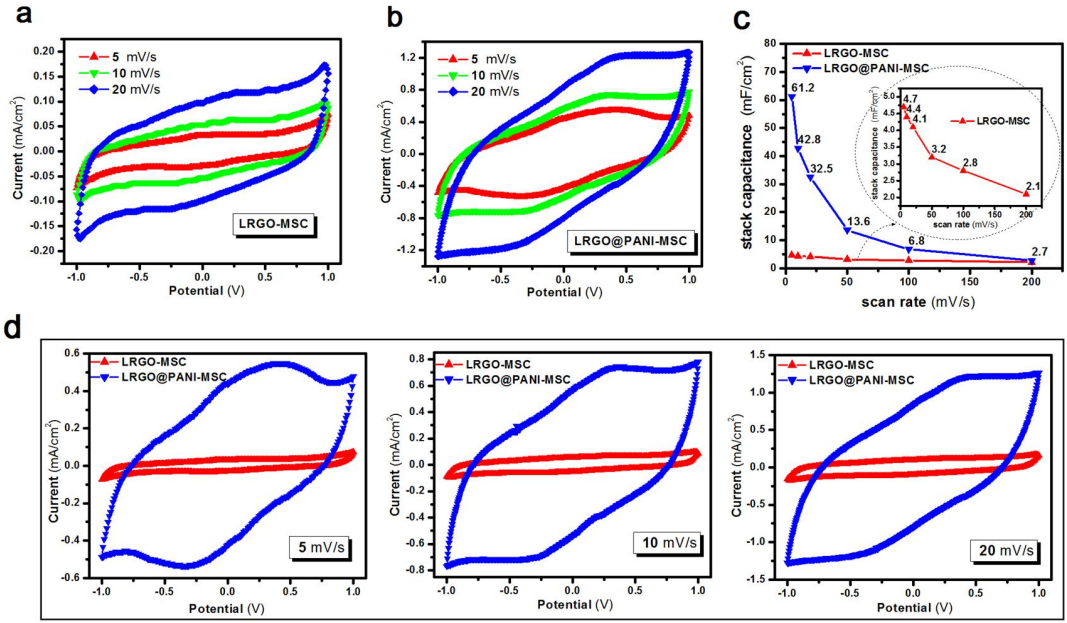

**Fig. 5: cyclic voltammetry of MSCs in PVA/H₂SO₄ electrolyte.** The CV curves of (**a**) LRGO-MSC and (**b**) LRGO@PANI-MSC show nearly rectangular shapes without chemical reaction peaks and a pair of redox peaks, respectively. (**c**) The stack area capacitance of MSCs as a function of the scan rate. (**d**). A comparison of CV curves at the scan rate of 5, 10 and 20 mV/s for LRGO-MSC and LRGO@PANI-MSC.

The capacitance of the micro-supercapacitor ($C_{stack}$) is calculated based on the area and volume of the stack. This includes the whole area (or volume) of the active material, current collector and the gap between the electrodes. The stack area capacitance values of the both MSCs as a function of the scan rate are shown in Fig. 5c and calculated by integrating the area within the CV curves using the equation[10,13]:

$$C_{stack} = \frac{\int I\, dV}{2k(\Delta V)\, A} \qquad (1)$$

where the numerator is the area under the CV curve, $\Delta V$ is the operation voltage window, $k$ is the scan rate and $A$ is the total area of supercapacitor. In Fig. 5c, the stack capacitance values are labeled at each scan rate for both samples. When the scan rate is 5 mV/s, the stack capacitance of the LRGO@PANI-MSC is as high as 61.2 mF/cm$^2$, which is about 13 times higher than that of LRGO-MSC. Even at a relatively fast scan rate of 50mV/s, the stack capacitance of an LRGO@PANI micro supercapacitor is about 5 times higher than the LRGO micro-supercapacitor's capacitance. It is observed that in FE-SEM images, Fig. 2, the LRGO@PANI micro electrodes exhibit the larger specific surface area, thinner graphene flakes and more porous than LRGO micro electrode's, all of which help to minimize the pathway for fast penetration of the electrolyte to the surface micro electrodes and fast charging-discharging reaction. In addition, irradiating graphene oxide by laser beam causes local reduction and formation of graphene and in LRGO@PANI sample, laser breaks oxygen-hydrogen (O-H) bond on graphene oxide sheets and replaces them by π-nitrogen bond of polyaniline chains; accordingly, formation of the electronic coupling large-scale π-π conjugation between PANI and LRGO facilitated charge transfer and electrical conductivity and consequently improved electrochemical performance significantly.

It is seen that as the scan rates increase, the expected capacitance would decrease as a result of limited electrolyte ions diffusion within the pores and depth of electrode active material, which shrink the electrochemical effective surface area and the cations/anions mobility through the active material. Actually, at higher scan rates, electrolytes ions ($H^+$ and $SO_4^{-2}$) touch outer surface of the electrode materials while at lower scan rates, they have enough time to get into deep porous material thus bring about higher capacitance. For comparison, the CV curves of the LRGO-MSC and LRGO@PANI-MSC cells are shown together at the scan rate of 5, 10 and 20 mV/s in Fig. 5d. It is observed that the current densities associated without polyanilline devices are markedly lower (approximately 6 orders of magnitude) than their reduced graphene/polyaniline counterparts, because of low porous structure and non-existent or very limited electroactive site in LRGO micro electrodes. Evidently, the peak current of LRGO@PANI electrodes is higher than that of pure LRGO electrode, indicating that the PANI could effectively increase the capacitance of RGO.

Fig. 6 illustrates the galvanostatic charge/discharge curves at various current density for LRGO-MSC and LRGO@PANI-MSC cells. The nearly symmetrical and triangular shape of galvanostatic charge/discharge (GCD) curves in Fig. 6 confirms a high reversibility of capacitance behavior of high-performance supercapacitor micro electrodes. This is in good agreement with the previous CV results. Fig. 6a demonstrates the GCD curves of LRGO-MSC device. No faradic interactions is observed which means that the capacitance is mainly attributed to pure electric double layer capacitances. But the GCD curves of LRGO@PANI electrodes show two potential stage in the ranges of 1.0-0.5 V and 0.5- (-1.0) V (Fig. 6b). The former stage with a relatively short discharging duration is attributed to electric double layer capacitances in interdigital fingers, however the latter stage with a much longer discharging duration is related to the synergistic effect between double layer capacitance and pseudocapacitance, corresponding to RGO and RGO@PANI that is in good agreement with others graphene/polyaniline composite based supercapasitors[25, 26]. By increasing current density, the charge/discharge process goes faster and performs from 375s at 0.07 mA/cm$^2$ to 18s at 0.7 mA/cm$^2$ for LRGO-MSC and 4200s at 0.07 mA/cm$^2$ to 240s at 0.7 mA/cm$^2$ for LRGO@PANI-MSC. The charge/discharge time in the GCD curves of LRGO@PANI electrode is enhanced by more 10 orders of magnitudes in comparison with LRGO electrode. The overwhelming long charge/discharge time is a result of the pseduocapactive characteristics of the LRGO@PANI electrode and a good charge storage property upon the presence of PANI in the energy storage systems.

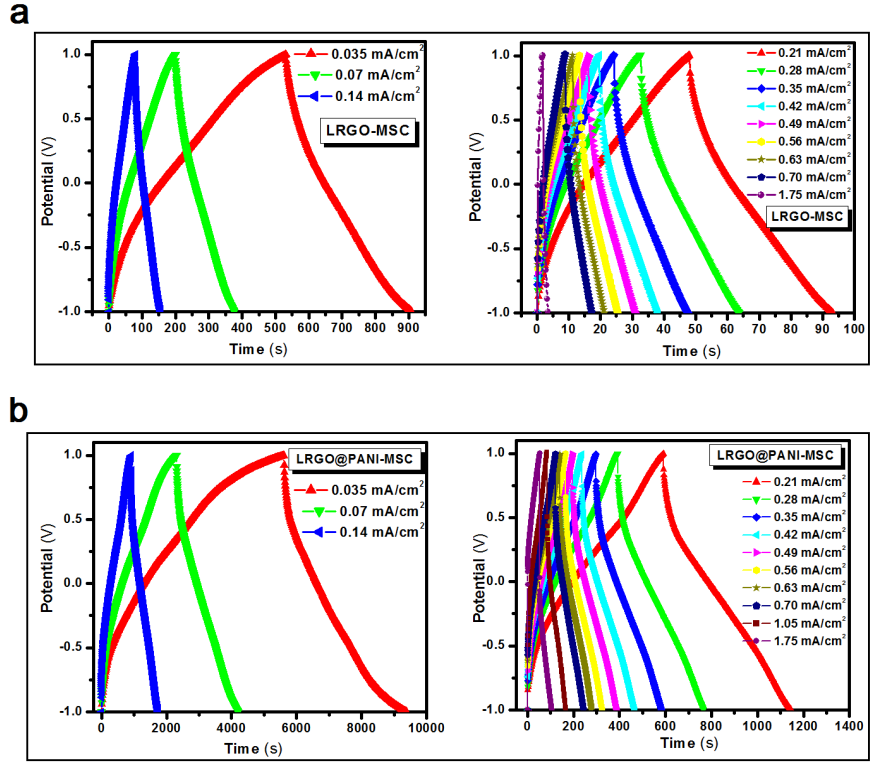

**Fig. 6: Galvanostatic charge/discharge curves of MSCs.** The GCD curves of (**a**) LRGO-MSC and (**b**) LRGO@PANI-MSC at different current density (from 0.035 to 1.75 mA/cm$^{-2}$).

The stack capacitance of LRGO@PANI-MSC is evaluated by galvanostatic charging/discharging at different current densities and compared with that of LRGO-MSC (Fig. 7a). The stack capacitance values, $C_{stack}$ (mF/cm$^2$) of the samples are estimated from the discharge process at the constant current density according to the following equation[10] and they are labeled by current density for both samples

$$C_{stack} = \frac{i \times t_D}{\Delta V \times A} \qquad (2)$$

where $i$ is the constant current (A), $t_D$ is discharge time (s), $\varDelta V$ is changings of potential during discharge process and $A$ is cell area. The stack capacitance 72.0 mF/cm² of the LRGO@PANI nanocomposite micro-electrodes at 0.035 mA/cm² is much higher compared to 6.6 mF/cm² of LRGO micro-electrodes. The stack capacitance of the LRGO@PANI-MSC retains 78.0% when the current density changed from 0.07 to 0.7 mA/cm² (it's 60% for LRGO-MSC), demonstrating the nanocomposites micro-electrodes with both high stack capacitance and good rate capability compared to LRGO micro-electrodes. This is probably due to the synergetic effect between the combination of electric double layer capacitances of RGO and faradaic capacitances of polyaniline component, which can shorten ion diffusion length.

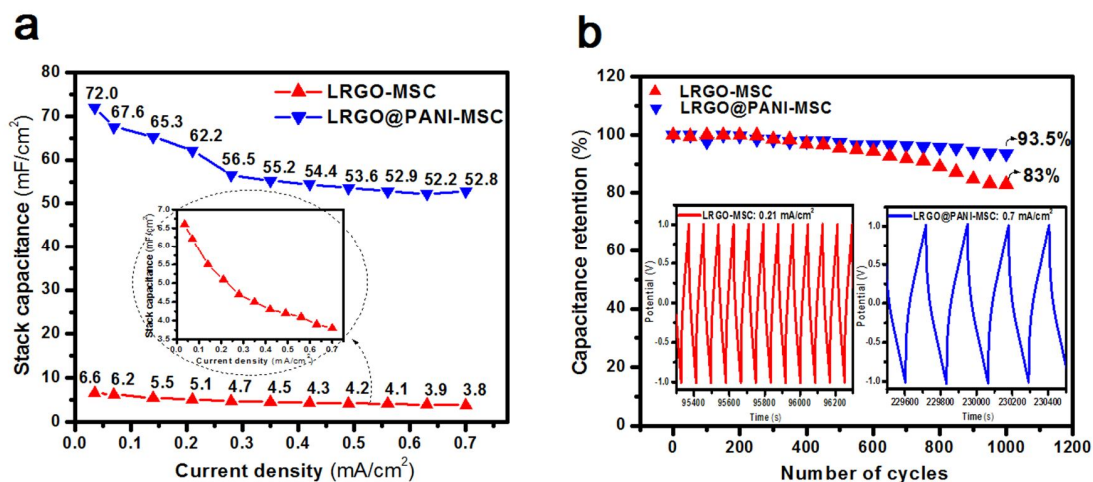

**Fig. 7: The stack area capacitance and life time of MSCs.** (**a**) The stack area capacitance values of LRGO and LRGO@PANI microsupercapacitor as calculated from the charge/discharge curves at different current densities. (**b**) Cycle life time shows excellent stability, losing about 6.5% of its initial area capacitance over 1,000 cycles for LRGO@PANI-MSC. It testes at 0.21 and 0.7 mA/cm² for LRGO-MSC and LRGO@PANI-MSC, respectively.

Long cycle life is a significant parameter for energy storage devices. Cycling stability of the fabricated micro supercapacitors are tested at 0.21 mA/cm² current density for LRGO-MSC and 0.7 mA/cm² current density for LRGO@PANI-MSC. The cycle life of charge/discharge is illustrated in Fig. 7b that shows more than 93.5% of the capacitance is retained after 1000 cycles for LRGO@PANI micro-electrode, which indicates excellent cycling stability of the nanocomposites micro-electrodes in long-term compared to LRGO micro-electrodes, 83%. This is due to the morphological and electrochemical changes of LRGO@PANI electrodes. This is confirmed by the results of GCD curves of different cycles that presented in Supplementary Fig. S5. The capacitance retention of micro supercapacitors in the present work is higher than previously reported RGO@PANI composite electrodes[27]. The insets of Fig. 7b illustrate the typical charge/discharge curve of both micro electrodes in the last 1000th cycle, exhibiting the stable symmetrical shape.

$iR$ drop ($V_{drop}$) at each constant current density (i) which is a measure of the overall resistance of the fabricated supercapasitor is obtained from following equation[13]:

$$R_{ESR} = \frac{V_{drop}}{2i} \qquad (3)$$

where $V_{drop}$ is the potential drop at the beginning of each discharge curve. The $iR$ drop for LRGO@PANI electrode, Fig. 6a, at different applied current densities is smaller than that of LRGO electrode, Fig. 6b. The $iR$ values of both LRGO and LRGO@PANI electrode at 0.14 mA/cm² are shown in Fig. 8a. It indicates that the inner resistance of the LRGO electrode is decreased with the presence of PANI chains grafted on RGO sheets, refer to FTIR results and Ref. 18. The pseudocapacitance of LRGO@PANI in the nanocomposite micro-electrodes are enhanced by the highly conductive polymer (polyaniline), which supports the faradic reactions of PANI component; hence, the supercapacitor based on LRGO@PANI nanocomposite has lower internal resistance. Lower internal resistance is of great importance in energy storage devices; for less energy will be wasted to produce unwanted heat during charging/discharging processes. Thus, LRGO@PANI micro electrodes are more suitable for fabricating safe and power saving supercapacitors compared with LRGO micro electrodes.

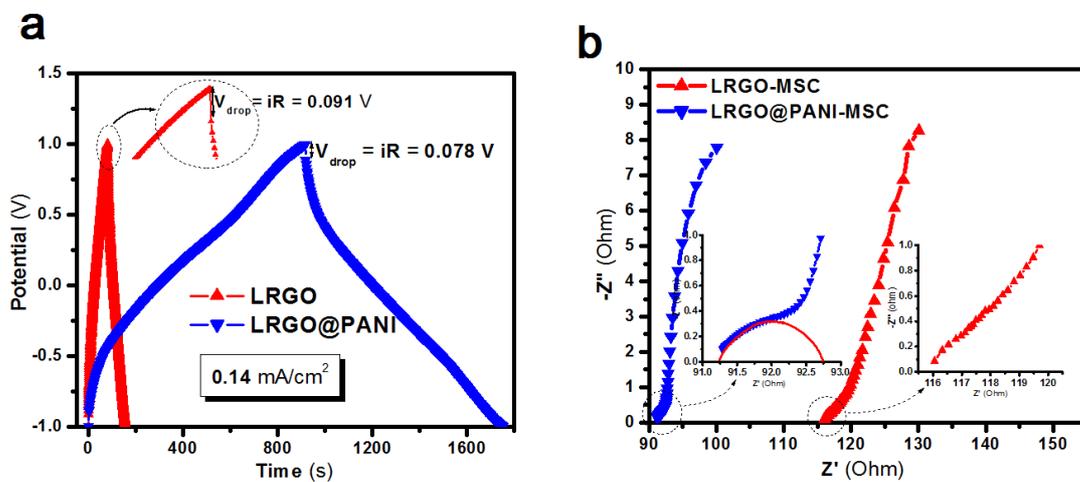

**Fig. 8: Potential drop and electrochemical impedance spectroscopy of MSCs. (a)** GCD curves and values of iR drop of MSCs at 0.14 mA/cm$^2$. **(b)** Nyquist plots of MSCs show an arc in the high frequency region for LRGO@PANI-MSC.

Furthermore, electrochemical impedance spectroscopy (EIS) of LRGO and LRGO@PANI electrodes, was performed; Nyquist plots of the LRGO-MSC and LRGO@PANI-MSC are shown in Fig. 8b, the inset represents the high-frequency range, and EIS measurements are recorded in frequency range of 0.1 Hz to 0.1 MHz. It is observed that the intercept between the impedance spectrum and real impedance axis (Z') is very small for two samples; which means that series resistance, including the electrolyte and contact resistance at the interface of active material/current collector, is very low. In the low frequency range for LRGO@PANI-MSC, the

Nyquist plot of samples show approximately straight vertical line close to an ideal supercapacitor characteristic. Vertical curve in comparison with that of the LRGO-MSC; reveals low ion/electrolyte diffusion resistance (or efficient ionic transfer between the electrodes and the electrolyte) while presenting PANI in the active materials. The impedance measurement of LRGO@PANI micro electrodes is linear in low-frequency region and an arc in high frequency region; the arc impedance plot is depended on charge transfer resistance at the electrode material/electrolyte interface, while the straight line indicates ion diffusion process. In high frequency region, an arc in the Nyquist plot of LRGO@PANI microelectrodes is associated to the electric resistance of PANI and ion transport to support the redox processes[28] The values of equivalent series resistances (ESR) are 115.8 and 92 Ω for the LRGO-MSC and LRGO@PANI-MSC, respectively. The decrease in the ESR of LRGO@PANI microelectrodes compared to that of LRGO microelectrodes is chiefly related to facile ion diffusion of the electrolyte inside the porous structure of composites electrode and better electrical conductivity. To demonstrate the overall performance of the micro supercapacitor cells, the volumetric energy, E (Wh/cm$^3$) and power densities, P (W/cm$^3$) of LRGO-MSC and LRGO@PANI-MSC based on the volume of the stack, are calculated at various charge/discharge current density using following equations[10]:

$$E = \frac{1}{2} \frac{C_{stack} \times (\Delta V)^2}{3600 \times h} \qquad (4)$$

$$P = \frac{E \times 3600}{t_D} \qquad (5)$$

where $C_{stack}$ is stack capacitance (mF/cm$^2$) and $\Delta V$ is the operating potential window (V), $h$ and $t_D$ are thickness (cm) of microelectrode and discharge time (s), respectively. Energy and power density values of the as prepared microsupercapacitor are given in Table 1 that the values are calculated at applied current density of 0.07 and 0.7 mA/cm$^2$. It is observed that the energy density of LRGO@PANI is more than 11 time that of LRGO-MSC. The energy density of LRGO@PANI with $H_2SO_4$/PVA gel electrolyte reaches 2.7 mWh/cm$^3$ at the power density of 81.4 mW/cm$^3$ which is higher than that of the other MSCs' reports based on graphene[29,30].

**Table 1. Energy and power density.** Calculated energy density (mWh/cm$^3$) and power density (mW/cm$^3$) for MSCs.

| Current density | 0.07 mA/cm$^2$ | | 0.7 mA/cm$^2$ | |
| --- | --- | --- | --- | --- |
| | Energy density (mWh/cm$^3$) | Power density (mW/cm$^3$) | Energy density (mWh/cm$^3$) | Power density (mW/cm$^3$) |
| **LRGO-MSC** | 0.47 | 9.5 | 0.18 | 78.6 |
| **LRGO@PANI-MSC** | 5.2 | 9.9 | 2.7 | 81.4 |

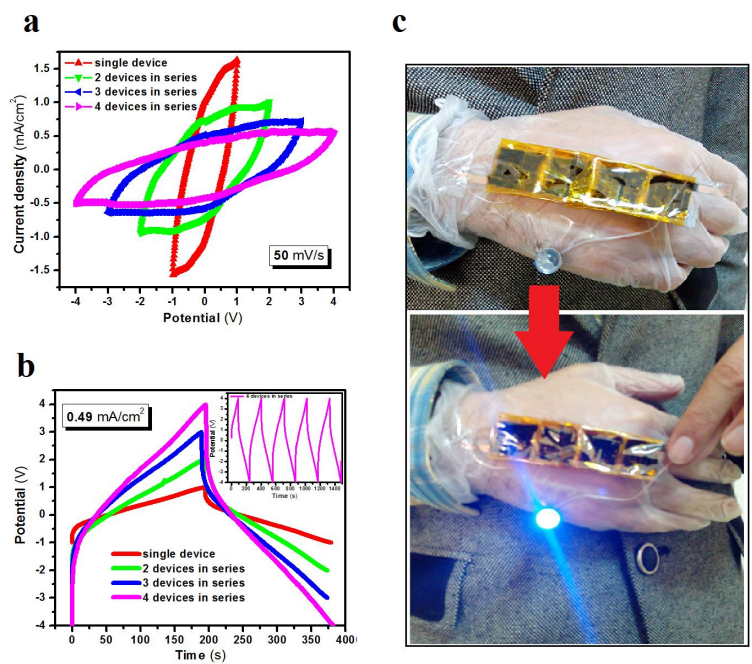

**Fig. 9: Electrochemical performance of LRGO@PANI-MSC in series combinations.** (**a**) CV and (**b**) GCD curves of LRGO@PANI-MSC for 1, 2, 3 and 4 cells in series (CV tests done at scan rate of 50 mV/s and GCD performed at current density of 0.49 mA/cm$^2$). (**c**) Photograph of a blue LED on glove powered by four flexible microsupercapacitors in series.

Since the energy that can be stored in a single micro supercapacitor is low for most practical usages thus, to demonstrate the possibility of applications of the LRGO@PANI-MSC, we investigated electrochemical performances of a tandem micro supercapacitor composed of four LRGO@PANI-MSC. As shown in Fig. 9a, b, the potential window of tandem LRGO@PANI-MSC exhibit very good electrochemical performance, enabling them to be applicable for electronic devices. Fig. 9c illustrates that the fully charged series of flexible micro supercapacitors can turn on a blue light-emitting diode (LED) on a glove with working voltage of 2.6 V.

## Discussion

Our results show that the best micro electrodes of supercapacitor is LRGO@PANI nanocomposite; owing to low internal resistance, very good charge transfers and excellent ionic conductivity and being very adhesive to substrates. Comparing to RGO; the LRGO@PANI based supercapacitors have the potential to serve as versatile flexible energy reservoir in novel future applications. Undergoing high cycling time spans with considerable capacitance retention promises long life utilization of LRGO@PANI micro electrode based devices. The prepared supercapacitor cells exhibit ideal symmetric capacitive behavior with nearly rectangular CV shape even in high potential window (at a high operating voltage of 2 V), that signifies electrolyte stability with no decomposition of the aqueous electrolyte into hydrogen or oxygen evolution. These results indicate the remarkable improvement of the extended operational voltage and capacitance that arises from the combination of the special architectural form of the LRGO@PANI electrodes with an aqueous electrolyte. Actually, Presence of PANI grafted on graphene sheets, not only could effectively inhibit the stacking/agglomerating of graphene, but also enhance electrode/electrolyte interface areas, improving high electrochemical performance of RGO@PANI nanocomposite. Besides, it is confirmed that the present approach is conducive to produce versatile graphene electrode structure. Laser reduction process here includes controlled photothermal reduction process and thus it renders less oxidative decomposition effect of GO. This method seems to pave the way for fabricating future flexible electronics and energy devices based on graphene nanocomposites by introducing a facile, economic and scalable method to design miscrosupercapacitor.

## Method

**Fabrication of LRGO@PANI-MSC.** LRGO@PANI-MSCs were prepared by laser beam through shadow mask. Steps for fabricating LRGO@PANI-MSC are as follow: first, high conducting Au patterns with in-plane interdigital geometry were coated on flexible PET substrate as current collectors by thermal evaporation method with the help of a designed mask with three digital fingers on each side (Figs. 1, 3(c1)). Then 2 ml of GO@PANI solution as active electrode materials of supercapacitor was drop casted on the interdigitated Au current collectors. GO@PANI solution is synthesized by aniline polymerization in aqueous dispersions of graphene oxide with 3.5mg/ml concentration, GO was synthesized via a modified Hummer's method[31]. Dried GO@PANI on PET were reduced in micro patterns by a homemade laser pointer (850 nm, 1200 mW Nichia laser diode) with the help of a designed shadow mask to form well adhesive RGO@PANI on the substrates. Finally, the as prepared symmetric interdigital micropatterns of LRGO@PANI were directly utilized to fabricate in-planar micro supercapacitor. The electrolyte ($H_2SO_4$/PAV, refer to supplementary information) was drop casted on the active electrode materials and left at room temperature for 2 h to evaporate extra water. Afterwards, the interdigitated active electrode materials are packed with Kapton tape to assemble micro supercapacitor cell and then the electrochemical test was performed.

**Fabrication of LRGO-MSC.** The interdigitated laser reduced graphene oxide (LRGO) micro electrodes were made in a similar method. The active electrode materials were prepared from 3.5mg/ml GO solution.

**Characterization and measurements.** The morphology of the as prepared microelectrodes is investigated using field emission scanning electron microscopy (TESCAN-MIRA3) and Raman (TESCAN-TAKRAM, 530-700 nm) and Fourier transform infrared (BRUCKER-TENSOR 27, 400-4000 $cm^{-1}$) spectroscopy are shown the structure properties of them. Electrochemical measurement of devices using a two-electrode setup is recorded by a potensiostat/galvanostat workstation (Ivium vertex) and EIS is performed on a Micro Autolab Type3 electrochemical workstations. All experiments and electrochemical analysis of devices are done in ambient atmosphere.

# References

1. Guozhen, S. & Zhiyong, F. (Eds.). *Flexible Electronics: From Materials to Devices*. World Scientific. (2016).

2.  Wang, R., Yao, M., & Niu, Z. Smart supercapacitors from materials to devices. *InfoMat*.1-13 (2019).

3.  Mannsfeld, S. C. et al. Highly sensitive flexible pressure sensors with microstructured rubber dielectric layers. *Nature materials*. **9**, 859 (2010).

4.  Chen, J. Supercapacitor-powered charger and implantable medical device. U.S. Patent Application 16/059,508, filed March 7, (2019).

5.  Cai, J., Chao, L. & Akira, W. Laser direct writing of high-performance flexible all-solid-state carbon micro-supercapacitors for an on-chip self-powered photodetection system. *Nano Energy* **30**, 790-800 (2016).

6.  He, Y. et al. Nano-sandwiched metal hexacyanoferrate/graphene hybrid thin films for in-plane asymmetric micro-supercapacitors with ultrahigh energy density. *Materials Horizons* **6**, 1041-1049 (2019).

7.  Ji, J. et al. Phytic acid assisted fabrication of graphene/polyaniline composite hydrogels for high-capacitance supercapacitors. *Composites Part B: Engineering* **155** 132-137 (2018).

8.  Frackowiak, E. & Béguin, F. Supercapacitors: Materials, Systems and Applications. (2013).

9.  Bavane, R. G. Synthesis and characterization of thin films of conducting polymers for gas sensing applications. (2014).

10. Ye, J. et al. Direct Laser Writing of Graphene Made from Chemical Vapor Deposition for Flexible, Integratable Micro-Supercapacitors with Ultrahigh Power Output. *Advanced Materials* **30**, 1801384 (2018).

11. Zhang, K. et al. Graphene/polyaniline nanofiber composites as supercapacitor electrodes. *Chemistry of Materials* **22**, 1392-1401 (2010).

12. Wang, F. et al. Rapid and low-cost laser synthesis of hierarchically porous graphene materials as high-performance electrodes for supercapacitors. *Journal of materials science* **54**, 5658-5670 (2019).

13. El-Kady, M. F. & Richard B. K. Scalable fabrication of high-power graphene micro-supercapacitors for flexible and on-chip energy storage. *Nature communications* **4**, 1475 (2013).

14. Zang, X. et al. Graphene/polyaniline woven fabric composite films as flexible supercapacitor electrodes. *Nanoscale* **7**, 7318-7322 (2015).

15. Wang, H. et al. Effect of graphene oxide on the properties of its composite with polyaniline. *ACS applied materials & interfaces* **2**, 821-828 (2010).

16. Li, D. et al. Processable aqueous dispersions of graphene nanosheets. *Nature nanotechnology* **3**, 101 (2008).

17. Chen, F. et al. Synthesis and microwave absorption properties of graphene-oxide (GO)/polyaniline nanocomposite with gold nanoparticles. *Chinese Physics B* **24**, 087801 (2015).

18. Vinoth, R., et al. Ruthenium based metallopolymer grafted reduced graphene oxide as a new hybrid solar light harvester in polymer solar cells. *Scientific reports*, **7**, 1-14(2017).

19. Wu, J. B. et al. Raman spectroscopy of graphene-based materials and its applications in related devices. *Chemical Society Reviews* **47**, 1822-1873 (2018).

20. Beams, R., Luiz, G. C. & Lukas, N. Raman characterization of defects and dopants in graphene. *Journal of Physics: Condensed Matter* **27**, 083002 (2015).

21. Luo, Y. et al. Self-assembled graphene@ PANI nanoworm composites with enhanced supercapacitor performance. *Rsc Advances* **3**, 5851-5859 (2013).

22. Fesenko, O. et al. Graphene-enhanced Raman spectroscopy of thymine adsorbed on single-layer graphene. *Nanoscale research letters* **10**, 163 (2015).

23. Stoller, M. D. & Rodney S. Ruoff. Best practice methods for determining an electrode material's performance for ultracapacitors." *Energy & Environmental Science* **3**, 1294-1301 (2010).

24. Zhang, Ch. et al. Planar integration of flexible micro-supercapacitors with ultrafast charge and discharge based on interdigital nanoporous gold electrodes on a chip. *Journal of Materials Chemistry A* **4**, 9502-9510 (2016).

25. Wu, Q. et al. Supercapacitors based on flexible graphene/polyaniline nanofiber composite films. *ACS nano* **4**, 1963-1970 (2010).

26. Mondal, S., Utpal, R. & Sudip, M. Reduced graphene oxide/Fe3O4/polyaniline nanostructures as electrode materials for an all-solid-state hybrid supercapacitor."*The Journal of Physical Chemistry C* **121**, 7573-7583 (2017).

27. Wang, Z. et al. Three-dimensional printing of polyaniline/reduced graphene oxide composite for high-performance planar supercapacitor. *ACS applied materials & interfaces* **10**, 10437-10444 (2018).

28. Sawangphruk, M. et al. High-performance supercapacitors based on silver nanoparticle–polyaniline–graphene nanocomposites coated on flexible carbon fiber paper. *Journal of Materials Chemistry A* **1**, 9630-9636 (2013).

29. Liu, W. W. et al. Superior micro-supercapacitors based on graphene quantum dots. *Advanced Functional Materials* **23**, 4111-4122 (2013).

30. Li, L. et al. High-performance pseudocapacitive microsupercapacitors from laser-induced graphene. *Advanced Materials* **28,** 838-845 (2016).

31. William, S., Hummers, J. R. & Richard, E. O. Preparation of graphitic oxide. *J. Am. Chem. Soc* **80**, 1339-1339 (1958).